\documentclass[aps,prl,reprint,superscriptaddress,showpacs]{revtex4-1}
\usepackage{graphicx}
\usepackage[dvips]{color}
\usepackage{multirow} \usepackage{rotating}
\usepackage{dcolumn}
\usepackage{bm}

\begin{document}

\preprint{Graphene V9}

\title{Anomalously strong pinning of the filling factor $\nu=2$ in epitaxial graphene}

\author{T.~J.~B.~M.~Janssen}
 \email{jt.janssen@npl.co.uk}

\author{A.~Tzalenchuk}%
\affiliation{
 National Physical Laboratory,\\ Hampton Road, Teddington, TW11 0LW, UK}

\author{R.~Yakimova}
\affiliation{Department of Physics, Chemistry and Biology (IFM), Link\"{o}ping University, S-581 83 Link\"{o}ping, Sweden}

\author{S.~Kubatkin}
\author{S. Lara-Avila}
\affiliation{Department of Microtechnology and Nanoscience, Chalmers University of Technology, S-412 96 G\"{o}tenborg, Sweden}

\author{S.~Kopylov}
\author{V.~I.~Fal'ko}
\affiliation{Physics Department, Lancaster University, Lancaster LA1 4YB, UK}

\date{\today}

\begin{abstract}

We explore the robust quantization of the Hall resistance in epitaxial graphene grown on Si-terminated SiC.
Uniquely to this system, the dominance of quantum over classical capacitance in the charge transfer between the substrate and graphene is such that Landau levels (in particular, the one at exactly zero energy) remain completely filled over an extraordinarily
broad range of magnetic fields. One important implication of this pinning of the filling factor is that
the system can sustain a very high nondissipative current. This makes epitaxial graphene ideally suited for quantum resistance metrology, and we have achieved a precision of 3 parts in $10^{10}$ in the Hall resistance quantization measurements.

\end{abstract}

\maketitle

The quantum Hall effect (QHE) is one of the key fundamental phenomena in solid-state physics \cite{Klitzing80}.
It was observed in two-dimensional electron systems in semiconductor materials and, since recently, in graphene:
both in exfoliated  \cite{Novoselov05,Zhang05,Castro-Neto09} and epitaxial  \cite{Shen09,Wu09,Tzalenchuk10,Jobst10,Tanabe10} devices.
A direct high-accuracy comparison of the conventional QHE in semiconductors with that observed in graphene constitutes
a test of the universality of this effect. The affirmative result would strongly support the pending
redefinition of the SI units based on the Planck constant $h$ and the electron charge $e$ \cite{Mills06}
and provide an international resistance standard based upon quantum physics \cite{Jeckelmann01}.

Graphene is believed to offer an excellent platform for QHE metrology due to the large energy separation between Landau levels (LL)
resulting from the Dirac-type ``massless'' electrons specific for its band structure \cite{Geim09}. The Hall resistance quantization
with an accuracy of 3 parts in $10^9$ has already been established \cite{Tzalenchuk10} in Hall-bar devices manufactured from epitaxial graphene  grown on Si-terminated face of SiC (SiC/G).
However, for graphene to be practically employed as an embodiment of a quantum resistance standard,
it needs to satisfy further stringent requirements \cite{Jeckelmann01}, in particular with respect
to \textit{robustness} over a range of temperature, magnetic field and measurement current.
A high measurement current, which a device can sustain at a given temperature without dissipation,
is particularly important for precision metrology as it defines the maximum attainable signal-to-noise ratio.

The extent of the QHE plateaux in conventional 2D electron systems is, usually, set by disorder and temperature.
Disorder pins the Fermi energy in the mobility gap of the 2D system, which suppresses
dissipative transport at low temperatures over a finite range of magnetic fields
around the values corresponding to exactly filled LLs.
These values can be calculated from the carrier density $n_s$ determined from the low-field Hall resistivity measurements
and coincide with the maximum non-dissipative current, the breakdown current.
Thus, the breakdown current in conventional two-dimensional semicondutors peaks very close to the field values
where the filling factor $\nu$ is an even integer \cite{Jeckelmann01}.
Though less studied experimentally, the behaviour of the breakdown current on the plateaux for the exfoliated graphene,
including the $\nu=2$ plateau corresponding to the topologically protected $N=0$ LL, looks quite similar \cite{Bennaceur10}.

In this Brief Report we explore the robustness of the Hall resistance quantization in SiC/G.
Unlike the QHE in conventional 2D systems, where the carrier density is independent of magnetic field, here specifically to SiC/G,
we find that the carrier density in graphene varies with magnetic field due to the charge transfer between surface donor states in SiC and graphene.
Most importantly,  we find magnetic field intervals of several Tesla, where the carrier density in graphene increases linearly with the magnetic field, resulting in the pinning of $\nu=2$ state with electrons at the the chemical potential occupying SiC surface donor states half-way between the $N=0$ and $N=1$ LLs in graphene.
Interestingly, at magnetic fields above the $\nu=2$ filling factor pinning interval, the carrier density saturates at a value up to 30\% higher than the zero-field carrier density.
The pinned filling factor manifests itself in a continuously increasing breakdown current toward the upper magnetic field end
of the $\nu=2$ state far beyond the nominal value of $B_{\nu=2}$ calculated from the zero-field carrier density.
Facilitated by the high breakdown current in excess of 500 $\mu \rm A$ at 14 T we have achieved a precision of 3 parts in $10^{10}$
in the Hall resistance quantization measurements.

The anomalous pinning of $\nu=4N+2$ filling factors in SiC/G is determined by the dominance of the quantum capacitance, $c_q$, \cite{Luryi88} over the classical capacitance per unit area, $c_c$, in the charge transfer between graphene and surface-donor states  of SiC/G: $c_q \ll c_c$, where $c_q=e^2\gamma_e$, $c_c=1/(4\pi d)$ and $\gamma_e$ is the density of states of electrons at the Fermi level. The latter reside in the 'dead layer' of carbon atoms, just underneath graphene \cite{vanBommel75,Riedl07,Varchon07,Mattausch07,Emtsev08,Qi10}.
This layer is characterised by a $6\sqrt{3}\times 6\sqrt{3}$ supercell of the reconstructed surface of sublimated SiC.
Missing or substituted carbon atoms in various positions of such a huge supercell in the dead layer create localised surface states with a broad distribution of energies within the bandgap of SiC ($\approx 2.4\ \rm eV$).

It appears that the density of such defects is higher in material grown at low temperatures ($1200-1600\rm ^\circ C$) resulting in graphene  doped to a large electron density, $n_s\sim10^{13}\rm\ cm^{-2}$, which is difficult to change \cite{Kopylov10}.
On the other hand, growth at higher temperatures, $T \approx 2000\rm ^\circ C$, and in a highly pressurised atmosphere of Ar seems to improve the integrity of the reconstructed 'dead' layer, leading to a lower density of donors on the surface and, therefore, producing graphene with a much lower initial doping \cite{Tzalenchuk10,Lara-Avila11}.

\begin{figure}
\includegraphics{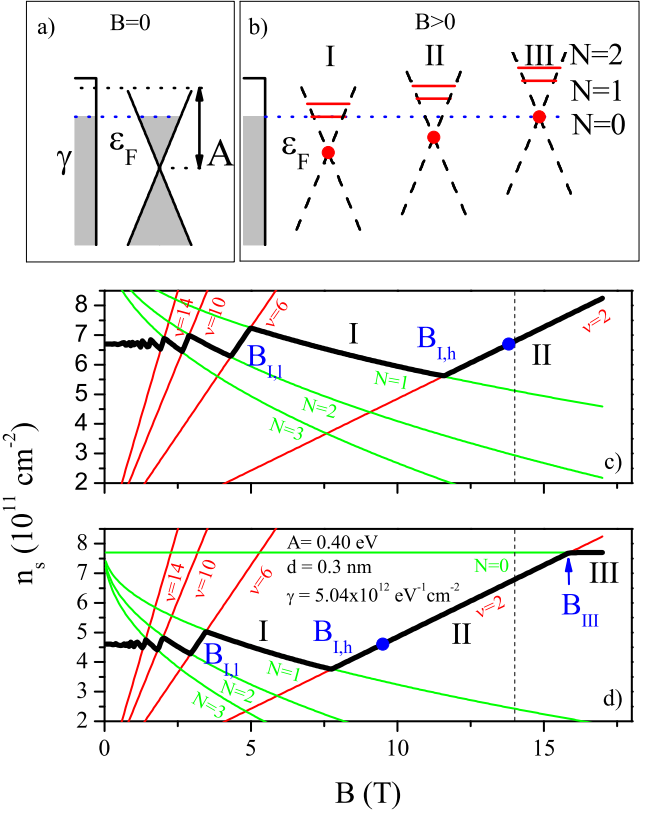}
\caption{\label{fig1} (Color online) Schematic band-structure for graphene on SiC in zero field (a) and in quantising field (b); graphical solution for carrier density as a function of magnetic field, $n_s(B)$,
of the charge-transfer model given by Eq.~\ref{eq1} (black line) together with lines
of constant filling factor (red/gray lines) and $n_s(B, N)$ (green/light lines) for $n_g=5.4\times 10^{11}\rm\ cm^{-2}$ (c) and $n_g=8.1\times 10^{11}\rm\ cm^{-2}$ (d). The vertical dashed line indicates the maximum field of 14 T in our setup and the blue dot indicates $\nu=2$ calculated from $n_s(0)$.}
\end{figure}

The quantum capacitance of a two-dimensional electron system is the result of a low compressibility of the
electron liquid determined by the peaks in $\gamma_e$.
For electrons in high-mobility GaAs/AlGaAs heterostructures in magnetic field, the quantum capacitance
manifests itself in weak magneto-oscillations of the electron density \cite{Eisenstein94,John04} due to the
suppressed density of states inside the inter-Landau level gaps.
A similarly weak effect has been observed in graphene exfoliated onto n-Si/SiO$_2$ substrate
\cite{Ponomarenko10}, where the influence of a larger (than in usual semiconductors) inter-LL gaps is hindered by a strong charging effect determined by a relatively large thickness of SiO$_2$ layer.
For epitaxial graphene on SiC, due to the short distance, $d\approx 0.3-0.4$~nm, between
the dead layer hosting the donors and graphene, the effect of quantum capacitance is much stronger, and the oscillations of electron density take the form of the robust pinning of the electron filling factor.
A similar behaviour was observed in STM spectroscopy of turbostratic graphite, where charge is transferred between the top graphene layer and the underlying bulk layers \cite{Song10}. The charge transfer in SiC/G is illustrated in the sketches in Fig.~\ref{fig1}, for $B=0$ (a) and quantising magnetic fields (b). The transfer can be described using the charge balance equation \cite{Kopylov10}
\begin{equation}\label{eq1}
\gamma [ A-4\pi e^{2}d(n_s+n_{g})-\varepsilon_{F}]=n_s+n_g.
\end{equation}
The left hand side of this equation accounts for the depletion of the surface donor states,
where $A$ is the difference between the work function of undoped graphene and the work function
of electrons in the surface donors in SiC, $\varepsilon_F$ is the Fermi energy of electrons in graphene,
and $\gamma$ is the density of donor states in the dead layer. An amount, $n_s$, of this charge density is transferred to graphene, and an amount, $n_g$, (controlled by the gate voltage) - to the polymer gate \cite{Lara-Avila11}.

Graphical solutions for the charge transfer problem for two values of $n_g$ are shown in Fig.~\ref{fig1} (c) and ~\ref{fig1}(d) for a broad range of magnetic fields. For graphene within interval III (visible only in the case of the higher $n_g$)
the Fermi energy coincides with the partially filled zero-energy LL, $\varepsilon_{F}=0$,
which determines the carrier density $n_{\infty}=\frac{A\gamma}{1+e^2\gamma/c_c}-n_g$,
and can be up to 30\% higher than the zero-field density $n_s(0)$ in the same device \cite{Kopylov10}.
This regime of fixed electron density is terminated at the low field end, at $B_{\rm III}=h n_{\infty} / 2e$, where
the $N=0$ LL is completely occupied by electrons with the density $n_{\infty}$. Note that for the $n_g$ presented here, $B_{\rm III}>14\rm\ T$ -- the maximum field in our setup.
Similarly, for magnetic field interval I, the Fermi level $\varepsilon_F=\hbar v \sqrt{2}/\lambda_B$
coincides with the partially filled $N=1$ LL ($\lambda_B=\sqrt{\hbar/eB}$), and, for this interval,
$n_s^{\rm I}=n_s(B, 1)$ with $n_s(B, N)=n_{\infty}-\frac{\gamma \hbar v \sqrt{2N}/\lambda_B}{1+e^2\gamma/c_c}$.
The interval I is limited by the field values for which the $N=1$ LL in the electron gas with the density $n_s^{\rm I}$ is emptied at the higher field end,
$B_{\rm I,h} = \frac{h}{2e}[\sqrt{n_{\infty}+\frac{\pi}{2}\frac{\gamma^2 v^2 \hbar^2}{(1+e^2\gamma/c_c)^2}} -
\sqrt{\frac{\pi}{2}}\frac{\gamma v \hbar}{1+e^2\gamma/c_c}]^2$,
and is full at the lower end,
$B_{\rm I,l} = \frac{h}{6e}[\sqrt{n_{\infty}+\frac{\pi}{6}\frac{\gamma^2 v^2 \hbar^2}{(1+e^2\gamma/c_c)^2}} -
\sqrt{\frac{\pi}{6}}\frac{\gamma v \hbar}{1+e^2\gamma/c_c}]^2$.
In magnetic field interval II the chemical potential in the system lies inside the gap between $N=0$ and $N=1$ LL in graphene. As a result, over this entire interval the $N=0$ LL in graphene is full and $N=1$ is empty, so that the filling factor in graphene is fixed at the value $\nu =2$, and the carrier density increases linearly with the magnetic field, $n_s = 2eB/h$, due to the charge transfer from SiC surface.

According to Eq. \ref{eq1}, lowering the carrier density using an electrostatic gate is equivalent to effectively
reducing the work function difference between graphene and donor states by $n_g(1/\gamma+e^2/c_c)$,
which shifts the range of the magnetic fields where pinning of the $\nu=2$ state takes
place. For instance, reducing the zero-field carrier density from $n_s = 6.7\times 10^{11}\rm\ cm^{-2}$  [Fig.~\ref{fig1}(c)] to $n_s = 4.6\times 10^{11}\rm\ cm^{-2}$  [Fig.~\ref{fig1} (d)] moves interval II from $11.5 \rm\ T<B_{II}<21.6 \rm\ T$ down to $7.7 \rm\ T<B_{II}<15.9 \rm\ T$, almost entirely within the experimental range.


In order to verify the predictions of the theory regarding the pinning of the $\nu =2$ filling factor and its implications for the resistance metrology, we studied the QHE in a polymer-gated epitaxial graphene sample with Hall bar geometry of width $W=35\ \rm \mu m$ and length $L=160\ \rm \mu m$.
Graphene  was grown at $2000 ^\circ \rm C$ and 1 atm Ar gas pressure on the Si-terminated face of a semi-insulating 4H-SiC(0001) substrate.
The as-grown sample had the zero-field carrier density $n_s=1.1\times10^{12}\ \rm cm^{-2}$. Graphene was encapsulated in a polymer bilayer,
a spacer polymer followed by an active polymer able to generate acceptor levels under UV light. At room temperature electrons
diffuse from graphene through the spacer polymer layer and fill the acceptor levels in the top polymer layer.
Such a photo-chemical gate allowed non-volatile control over the charge carrier density in graphene. More fabrication details can
be found elsewhere \cite{Tzalenchuk10,Lara-Avila11}.

\begin{figure}
\includegraphics{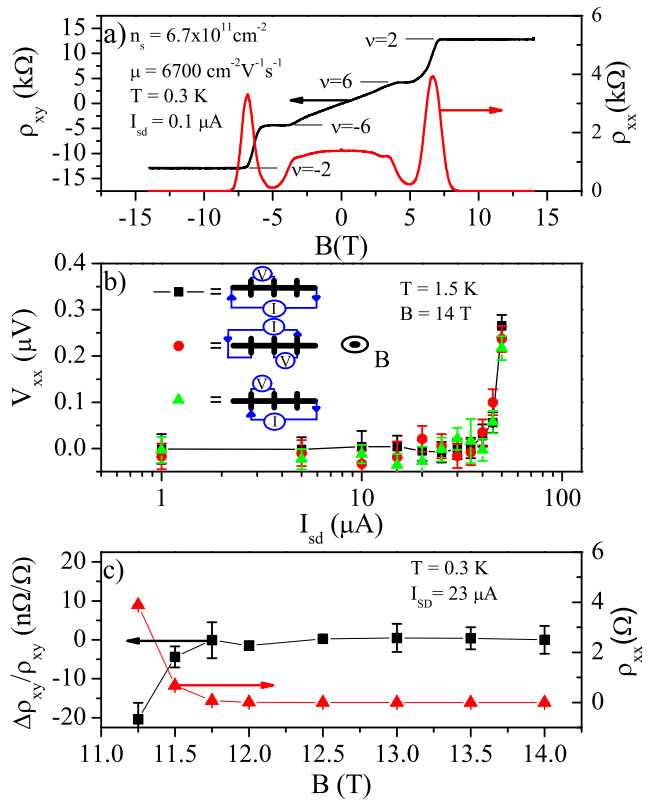}
\caption{\label{fig2}  (Color online) (a) Transverse ($\rho_{ xy}$) and longitudinal ($\rho_{xx}$) resistivity measurement. The horizontal lines indicate the exact quantum Hall resistivity values for filling factors $\nu = \pm 2$ and $\pm 6$. (b) Determination of the breakdown current, $I_{c}$, for three different measurement configurations explained in legend. (c) High-precision measurement of $\rho_{xy}$ and $\rho_{xx}$ as a function of magnetic field.
$\Delta\rho_{xy} /\rho_{xy}$ is defined as $(\rho_{xy}(B)- \rho_{xy}(14T))/\rho_{xy}(14T)$
and $\rho_{xy}(B)$ is measured relative to a $100\ \Omega$ standard resistor previously
calibrated against a GaAs quantum Hall sample \cite{Tzalenchuk10}. All error bars are $1\sigma$.}
\end{figure}

Figure \ref{fig2}(a) shows magneto transport measurements on the encapsulated sample tuned
to a zero-field carrier density of $n_s = 6.7\times 10^{11}\rm\ cm^{-2}$ corresponding to the case in fig.~\ref{fig1}(c)]. From the carrier density we estimate that the magnetic field $B_{\nu=2}$ needed for exact filling factor $\nu =2$ in this device is ~13.8~T.
A well-quantized Hall plateau in $\rho_{xy}$ can be seen at $\nu =\pm 2$ for both magnetic field directions which is more than 5~T wide, whereas the longitudinal resistivity, $\rho_{xx}$, drops to zero signifying a non-dissipative state.
In addition, a less precisely quantized plateau is present at $\nu = \pm 6$, for which $\rho_{xx}$ remains finite.

Accurate quantum Hall resistance measurements require that the longitudinal voltage remains zero
(in practice, below the noise level of the nanovolt meter) to ensure the device is in the non-dissipative state, which can be violated by the breakdown of the QHE at high current.
Figure~\ref{fig2}(b) shows the determination of the breakdown current $I_{c}$ at $B=14\rm\ T$ on the $\nu=2$ plateau.
Here we define $I_{c}$ as the source-drain current, $I_{sd}$, at which $V_{xx}\geq 10\ \rm nV$.
We find for three different combinations of source-drain current contacts that
the breakdown current for this value of $n_s$ is approximately $50\ \rm\mu A$ (note that $I_{sd}$ in a practical quantum Hall to $100\ \rm\Omega$
resistance measurement is $\rm \approx 25\ \mu A$ \cite{Williams10}).
The contact resistance, determined via a three-terminal measurement in the nondissipative state, is smaller than $\rm 1.5\ \Omega$.

Figure~\ref{fig2}(c) shows a precision measurement of $\rho_{xy}$ and $\rho_{xx}$ for different magnetic fields along the $\nu = 2$ plateau. Note that this plateau appears much shorter in the magnetic field range than that shown in Fig.~\ref{fig2} (a) because of the ~200 times larger measurement current used in precision measurements.
From this figure we determine that the mean of $\Delta\rho_{xy} /\rho_{xy}$ is $-0.06\pm 0.3\times10^{-9}$ for the data between 11.75 and 14.0~T,
while at the same time $\rho_{xx} < 1\ m\Omega$.
This result represents an order of magnitude improvement of QHE precision measurements in graphene,
as compared to the earlier record \cite{Tzalenchuk10}. Not only is QHE accurate, but it is also extremely robust in this epitaxial graphene device, easily meeting the stringent criteria for accurate quantum Hall resistance measurements normally applied to semiconductor systems.

\begin{figure}
\includegraphics{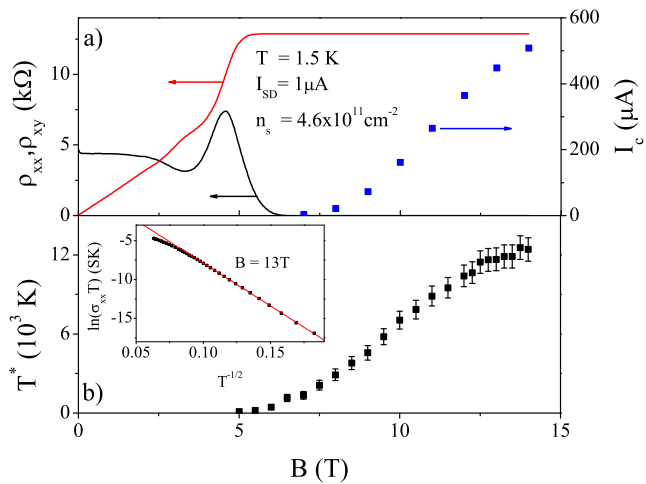}
\caption{\label{fig3} (Color online) (a) Experimental $\rho_{xx}$ (black line) and $\rho_{xy}$ (red/gray line)
together with the measured break-down current, $I_c$ (blue squares).
(b) Hopping temperature, $T^*$ as a function of magnetic field. Inset: $\ln(\sigma_{xx}T)$ versus $T^{-1/2}$ at 13~T.
Red/gray line is linear fit for $100>T>5\ \rm K$ giving $T^*\approx 12000\ \rm K$.}
\end{figure}

Using the polymer gating method \cite{Lara-Avila11}, we further reduce the zero-field electron density $n_s$ in graphene
to correspond to the solution of the charge transfer problem in Fig.~\ref{fig1}(d), i.e. down to $4.6\times 10^{11}\ \rm cm^{-2}$ as evidenced by magnetotransport measurements in  Fig.~\ref{fig3}(a).
On the $\nu=2$ quantum Hall resistance plateau we measure the breakdown current $I_c$, defined above, as a function of the magnetic field. Unlike the conventional QHE materials  \cite{Jeckelmann01}, the breakdown current in Fig.~\ref{fig3}(a) continuously increases from zero to
almost $500\ \mu\rm A$ far beyond $B_{\nu=2}\sim 9.5\ \rm T$ calculated from the zero-field carrier density. This is a direct consequence of the exchange of carriers between graphene and the donors in the 'dead' layer, which keeps the $N=0$ LL completely filled well past $B_{\nu=2}$.

The magnetic field range where the Fermi energy in SiC/G lies half-way between the $N=0$ and $1$ LLs
determines the activation energy $\hbar\sqrt{1/2} v/\lambda_B\sim 1000\ \rm K$ for the dissipative transport.
For such a high activation energy, the low-temperature dissipative transport is most likely to proceed
through the variable range hopping (VRH) between surface donors in SiC involving virtual occupancy
of the LL states in graphene to which they are weakly coupled. Indeed, as shown in the inset of Fig.~\ref{fig3}(b),
the temperature dependence of the conductivity $\sigma_{xx}$ measured at $B=13\ \rm T$ obeys
an $\exp(-\sqrt{T^*/T})$ dependence typical of the VRH mechanism. The $T^*$ values determined from the measurements
at different magnetic fields are plotted in the main panel of fig.~\ref{fig3}(b). The breakdown current rising
with field to very large values [Fig.~\ref{fig3}(a)] corresponds to  $T^*$ reaching extremely large values in excess of $10^4\ \rm K$ -- at least an order of magnitude larger than that observed in GaAs \cite{Furlan98} and more recently
in exfoliated graphene \cite{Giesbers09,Bennaceur10}.

In conclusion, we have studied the robust Hall resistance quantization in a large epitaxial graphene sample grown on SiC.
We have observed the pinning of the $\nu=2$ state which is consistent with our picture of magnetic field dependent charge transfer
between the SiC surface and graphene layer.
Together with the large break-down current this makes graphene on SiC the ideal system for high-precision resistance metrology.

\begin{acknowledgments}
We are grateful to T. Seyller, A. Geim, K. Novoselov and K. von Klitzing for discussions. This work was supported by the NPL Strategic Research programme, Swedish Research Council and Foundation for
Strategic Research, EU FP7 STREPs ConceptGraphene and SINGLE, EPSRC grant EP/G041954 and the Science \& Innovation Award EP/G014787.
\end{acknowledgments}

\nocite{*}

\providecommand{\noopsort}[1]{}\providecommand{\singleletter}[1]{#1}%

\end{document}